\documentclass[10pt,a4paper,onecolumn]{IEEEtran}
\usepackage[lmargin=2cm, rmargin=2cm, tmargin=2cm, bmargin=2cm]{geometry}
\usepackage{ccaption}
\usepackage[T1]{fontenc}
\usepackage{graphicx}
\usepackage{fancyhdr}
\usepackage{helvet}
\usepackage{floatflt}
\usepackage{colortbl}
\usepackage{fancyvrb}
\usepackage{wrapfig}
\usepackage[round,authoryear]{natbib}
\usepackage[latin1]{inputenc}
\usepackage[T1]{fontenc}
\usepackage[english]{babel}
\usepackage{graphicx}
\usepackage{subfigure}
\usepackage{amsfonts}
\usepackage{amsmath}
\usepackage{theorem}
\usepackage{ae}

\newtheorem{assumption}{Assumption}
\newtheorem{remark}{Remark}
\newtheorem{theorem}{Theorem}

\title{An adaptive fuzzy dead-zone compensation scheme for nonlinear systems}
\author{Wallace Moreira Bessa, Max Suell Dutra, Edwin Kreuzer}
\date{}

\begin{document}

\maketitle

\abstract{

The dead-zone nonlinearity is frequently encountered in many industrial automation equipments and 
its presence can severely compromise control system performance. Due to the possibility to express 
human experience in an algorithmic manner, fuzzy logic has been largely employed in the last decades 
to both control and identification of uncertain dynamical systems. In spite of the simplicity of 
this heuristic approach, in some situations a more rigorous mathematical treatment of the problem 
is required. In this work, an adaptive fuzzy controller is proposed for nonlinear systems subject to 
dead-zone input. The convergence properties of the tracking error will be proven using Lyapunov 
stability theory and Barbalat's lemma. An application of this adaptive fuzzy scheme to a Van der 
Pol oscillator is introduced to illustrate the controller design method. Numerical results are 
also presented in order to demonstrate the control system performance.

}

\section{INTRODUCTION}

Dead-zone is a hard nonlinearity that can be commonly found in many industrial actuators, especially 
those containing hydraulic valves and electric motors. Dead-zone characteristics are often unknown 
and, as previously reported in the research literature, its presence can drastically reduce control 
system performance and lead to limit cycles in the closed-loop system.

The increasing number of works dealing with systems subject to dead-zone input shows the great 
interest of the engineering community in this particular nonlinearity. The most common approaches
are adaptive schemes \citep{tao1,wang1,zhou1,ibrir1}, fuzzy systems \citep{kim1,oh1,lewis1}, neural 
networks \citep{selmic1,tsai1,zhang1} and variable structure methods \citep{corradini1,shyu1}. Many 
of these works \citep{tao1,kim1,oh1,selmic1,tsai1,zhou1} use an inverse dead-zone to compensate the 
negative effects of the dead-zone nonlinearity even though this approach leads to a discontinuous 
control law and requires instantaneous switching, which in practice can not be accomplished with 
mechanical actuators. An alternative scheme, without using the dead-zone inverse, was originally 
proposed by \citet{lewis1} and also adopted by \citet{wang1}. In both works, the dead-zone is treated 
as a combination of a linear and a saturation function. This approach was further extended by 
\citet{ibrir1} and by \citet{zhang1}, in order to accommodate non-symmetric dead-zones.

Intelligent control, on the other hand, has proven to be a very attractive approach to cope with uncertain nonlinear systems 
\citep{tese, cobem2005,Bessa2017,Bessa2018,Bessa2019,Deodato2019,Lima2018,Lima2020,Lima2021,Tanaka2013}. 
By combining nonlinear control techniques, such as feedback linearization or sliding modes, with adaptive intelligent algorithms, 
for example fuzzy logic or artificial neural networks, the resulting intelligent control strategies can deal with the nonlinear 
characteristics as well as with modeling imprecisions and external disturbances that can arise.

This paper presents an adaptive fuzzy controller for nonlinear systems subject to dead-zone input. 
An unknown and non-symmetric dead-band is assumed. The dead-zone nonlinearity is also considered as 
a combination of linear and saturation functions, but an adaptive fuzzy inference system is introduced,
as universal function approximator, to cope with the unknown saturation function. Based on a Lyapunov-like 
analysis using Barbalat's lemma, the convergence properties of the closed-loop system is analytically proven. 
To show the applicability of the proposed control scheme, a Van der Pol oscillator is chosen as illustrative 
example. Simulation results of the adopted mechanical system are also presented to demonstrate the control 
system efficacy. 

\section{PROBLEM STATEMENT AND CONTROL SYSTEM DESIGN}

\begin{figure}[hbt]
\centering
\includegraphics[width=0.3\textwidth]{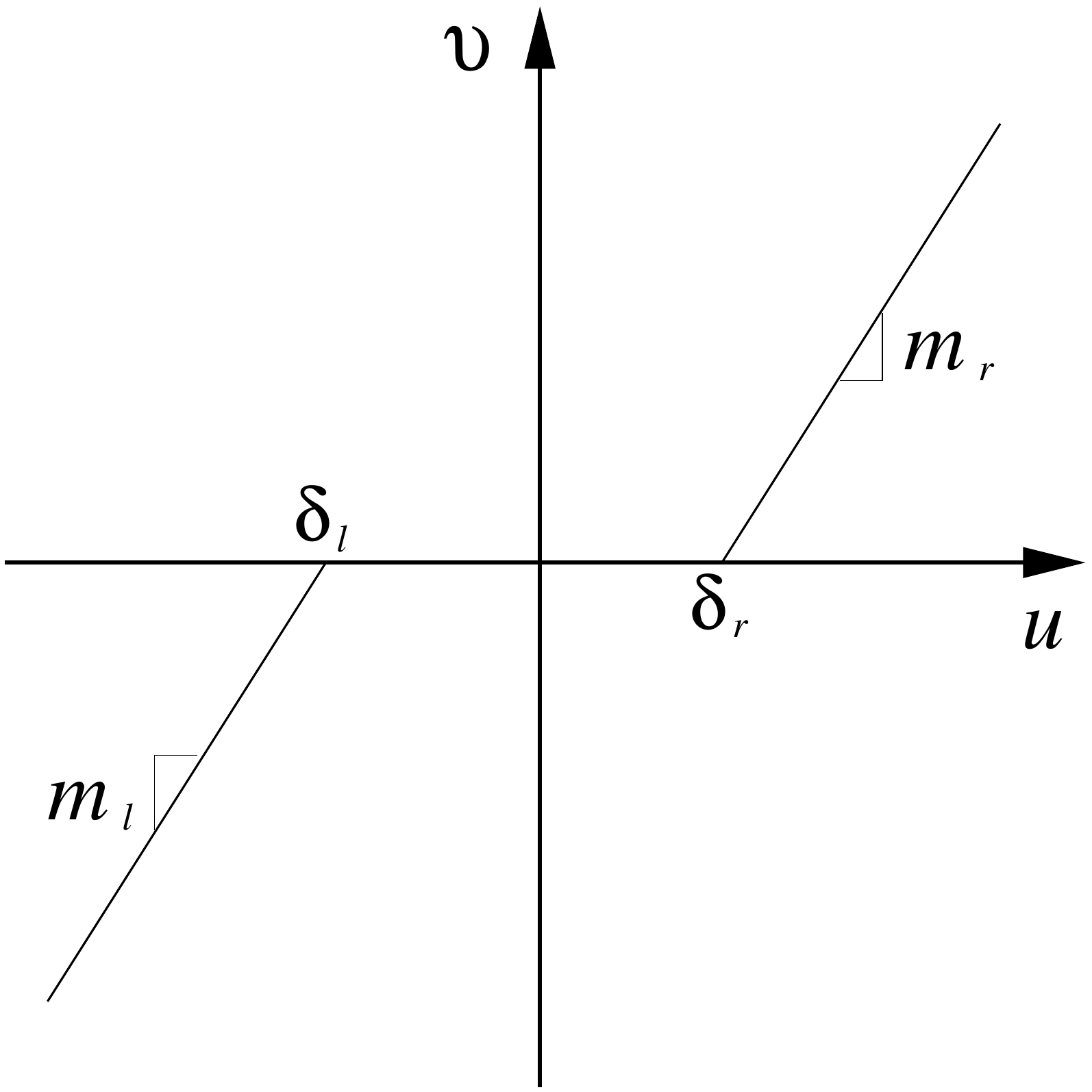} 
\caption{Dead-zone nonlinearity.}
\label{fig:dzone}
\end{figure}

Consider a class of $n^\mathrm{th}$-order nonlinear and non-autonomous systems:

\begin{equation}
x^{(n)}=f(\mathbf{x},t)+b(\mathbf{x},t)\upsilon
\label{eq:system}
\end{equation}

\noindent
where the scalar variable $x$ is the output of interest, $x^{(n)}$ is the $n$-th derivative of 
$x$ with respect to time $t$, $\mathbf{x}=[x,\dot{x},\ldots,x^{(n-1)}]$ is the system state vector, 
$f,b:\mathbb{R}^n\rightarrow\mathbb{R}$ are both nonlinear functions and $\upsilon$ represents the 
output of a dead-zone function, as shown in Fig.~\ref{fig:dzone}.

The dead-zone nonlinearity presented in Fig.~\ref{fig:dzone} can be mathematically described by:

\begin{equation}
\upsilon= \left\{\begin{array}{ll}
m_l\,(u-\delta_l)&\mbox{if}\quad u\le\delta_l \\
0&\mbox{if}\quad  \delta_l<u<\delta_r\\
m_r\,(u-\delta_r)&\mbox{if}\quad u\ge\delta_r
\end{array}\right.
\label{eq:dzone1}
\end{equation}

\noindent
where $u$ represents the controller output variable.

Considering that in many engineering components, as for instance hydraulic valves and electric motors, 
the slopes in both sides of the dead-zone are similar, the following physically motivated assumptions 
can be made for the dead-zone model presented in Eq.~(\ref{eq:dzone1}):

\begin{assumption}
The dead-zone output $\upsilon$ is not available to be measured.
\label{hp:output}
\end{assumption}
\begin{assumption}
The slopes in both sides of the dead-zone are equal and positive, i.e., $m_l=m_r=m>0$.
\label{hp:slope}
\end{assumption}
\begin{assumption}
The dead-band parameters $\delta_l$ and $\delta_r$ are unknown but bounded and with known signs, 
i.e., $\delta_{l\,\mathrm{min}}\le\delta_l\le\delta_{l\,\mathrm{max}}<0$ and $0<\delta_{r\,\mathrm{min}}
\le\delta_r\le\delta_{r\,\mathrm{max}}$.
\label{hp:band}
\end{assumption}

In this way, Eq.~(\ref{eq:dzone1}) can be rewritten in a more appropriate form \citep{lewis1,wang1}: 

\begin{equation}
\upsilon=m[u-d(u)] 
\label{eq:dzone2}
\end{equation}

\noindent
where $d(u)$ can be obtained from Eq.~(\ref{eq:dzone1}) and Eq.~(\ref{eq:dzone2}):

\begin{equation}
d(u)= \left\{\begin{array}{ll}
\delta_l&\mbox{if}\quad u\le\delta_l \\
u&\mbox{if}\quad  \delta_l<u<\delta_r\\
\delta_r&\mbox{if}\quad u\ge\delta_r
\end{array}\right.
\label{eq:dzone3}
\end{equation}

\begin{remark}
Considering Assumption~\ref{hp:band} and Eq.~(\ref{eq:dzone3}), it can be easily verified that $d(u)$ is 
bounded: $|d(u)|\le\delta$, where $\delta=\mathrm{max}\{-\delta_{l\,\mathrm{min}},\delta_{r\,\mathrm{max}}\}$.
\label{rm:dbounds}
\end{remark}

The proposed control problem is to ensure that, even in the presence of an unknown dead-zone input, 
the state vector $\mathbf{x}$ will follow a desired trajectory $\mathbf{x}_d=[x_d,\dot{x}_d,\ldots,
x^{(n-1)}_d]$ in the state space.

Regarding the development of the control law, the following assumptions should also be made:

\begin{assumption}
The state vector $\mathbf{x}$ is available.
\label{hp:stat}
\end{assumption}
\begin{assumption}
The desired trajectory $\mathbf{x}_d$ is once differentiable in time. Furthermore, every element of 
vector $\mathbf{x}_d$, as well as $x^{(n)}_d$, is available and with known bounds.
\label{hp:traj}
\end{assumption}

Let $\tilde{x}=x-x_d$ be defined as the tracking error in the variable $x$, and 

\begin{displaymath}
\mathbf{\tilde{x}}=\mathbf{x}-\mathbf{x}_d=[\tilde{x},\dot{\tilde{x}},\ldots,\tilde{x}^{(n-1)}]
\end{displaymath}

\noindent 
as the tracking error vector. 

Now, consider a combined tracking error measure: 

\begin{equation}
\varepsilon=\mathbf{c^\mathrm{T}\tilde{x}}
\label{eq:measu}
\end{equation}

\noindent
where $\mathbf{c}=[c_{n-1}\lambda^{n-1},\ldots,c_1\lambda,c_0]$ and $c_i$ states for binomial 
coefficients, i.e.,

\begin{equation}
c_i=\binom{n-1}{i}=\frac{(n-1)!}{(n-i-1)!\:i!}\:,\quad\quad i=0,1,\ldots,n-1 
\label{eq:binom}
\end{equation}

\noindent
which makes $c_{n-1}\lambda^{n-1}+\cdots+c_1\lambda+c_0$ a Hurwitz polynomial. 

From Eq.~(\ref{eq:binom}), it can be easily verified that $c_0=1$, for $\forall n\ge1$. Thus, for 
notational convenience, the time derivative of $\varepsilon$ will be written in the following form:

\begin{equation}
\dot{\varepsilon}=\mathbf{c^\mathrm{T}\dot{\tilde{x}}}
=\tilde{x}^{(n)}+\mathbf{\bar{c}^\mathrm{T}\tilde{x}}
\label{eq:dmeasu}
\end{equation}

\noindent
where $\mathbf{\bar{c}}=[c_{n-1}\lambda^{n-1},\ldots,c_1\lambda,0]$.

Based on Assumptions~\ref{hp:stat}~and~\ref{hp:traj}, the following control law can be proposed:

\begin{equation}
u=\frac{1}{bm}(-f+x^{(n)}_d-\mathbf{\bar{c}^\mathrm{T}\tilde{x}}-\kappa\varepsilon)+\hat{d}(\hat{u})
\label{eq:law}
\end{equation}

\noindent
where $\kappa$ is a strictly positive constant and $\hat{d}(\hat{u})$ an estimate of $d(u)$, that will 
be computed in terms of the equivalent control $\hat{u}=(bm)^{-1}(-f+x^{(n)}_d-\mathbf{\bar{c}^\mathrm{T}
\tilde{x}})$ by an adaptive fuzzy algorithm.

The adopted fuzzy inference system was the zero order TSK (Takagi--Sugeno--Kang), whose rules can be stated
in a linguistic manner as follows:

\begin{center}
\textit{If $\hat{u}$ is $\hat{U}_r$ then $\hat{d}=\hat{D}_r$}\hspace*{5pt};\hspace*{5pt}$r=1,2,\cdots,N$ 
\end{center}

\noindent
where $\hat{U}_r$ are fuzzy sets, whose membership functions could be properly chosen, and $\hat{D}_r$ is 
the output value of each one of the $N$ fuzzy rules.

Considering that each rule defines a numerical value as output $\hat{D}_r$, the final output $\hat{d}$ 
can be computed by a weighted average: 

\begin{equation}
\hat{d}(\hat{u}) = \frac{\sum_{r=1}^{N} w_r \cdot \: \hat{d}_r}{\sum_{r=1}^{N} w_r}
\label{eq:dc_media}
\end{equation}

\noindent
or, similarly,

\begin{equation}
\hat{d}(\hat{u}) = \mathbf{\hat{D}}^{\mathrm{T}}\mathbf{\Psi}(\hat{u})
\label{eq:dc_vetor}
\end{equation}

\noindent
where, $\mathbf{\hat{D}}=[\hat{D}_1,\hat{D}_2,\dots,\hat{D}_N]$ is the vector containing the attributed 
values $\hat{D}_r$ to each rule $r$, $\mathbf{\Psi}(\hat{u})=[\psi_1(\hat{u}),\psi_2(\hat{u}),\dots,$
$\psi_N(\hat{u})]$ is a vector with components $\psi_r(\hat{u})= w_r/\sum_{r=1}^{N}w_r$ and $w_r$ is 
the firing strength of each rule.

To ensure the best possible estimate $\hat{d}(\hat{u})$, the vector of adjustable parameters can be 
automatically updated by the following adaptation law:

\begin{equation}
\mathbf{\dot{\hat{D}}}=-\varphi\varepsilon\mathbf{\Psi}(\hat{u})
\label{eq:adapta}
\end{equation}

\noindent
where $\varphi$ is a strictly positive constant related to the adaptation rate. 

The boundedness and convergence properties of the closed-loop system are established in the following
theorem.

\begin{theorem}
\label{th:stab}
Consider the nonlinear system (\ref{eq:system}) subject to the dead-zone (\ref{eq:dzone1}) and 
Assumptions~\ref{hp:output}--\ref{hp:traj}. Then, the controller defined by (\ref{eq:law}), 
(\ref{eq:dc_vetor}) and (\ref{eq:adapta}) ensures the boundedness of all closed-loop signals and  
the exponential convergence of the tracking error, i.e., $\mathbf{\tilde{x}\to0}$ as $t\to\infty$.
\end{theorem}

\noindent
{\bf Proof:} Let a positive definite Lyapunov function candidate $V$ be defined as

\begin{equation}
V(t)=\frac{1}{2}\varepsilon^2+\frac{bm}{2\varphi}\mathbf{\Delta}^{\mathrm{T}}\mathbf{\Delta}
\label{eq:liap}
\end{equation}

\noindent
where $\mathbf{\Delta}=\mathbf{\hat{D}}-\mathbf{\hat{D}}^*$ and $\mathbf{\hat{D}}^*$ is the optimal
parameter vector, associated to the optimal estimate $\hat{d}^*(\hat{u})=d(u)$.

\noindent
Thus, the time derivative of $V$ is

\begin{align*}
\dot{V}(t)&=\varepsilon\dot{\varepsilon}+bm\varphi^{-1}\mathbf{\Delta^\mathrm{T}\dot{\Delta}}\\
&=(\tilde{x}^{(n)}+\mathbf{\bar{c}^\mathrm{T}\tilde{x}})\varepsilon
+bm\varphi^{-1}\mathbf{\Delta^\mathrm{T}\dot{\Delta}}\\
&=(x^{(n)}-x^{(n)}_d+\mathbf{\bar{c}^\mathrm{T}\tilde{x}})\varepsilon
+bm\varphi^{-1}\mathbf{\Delta^\mathrm{T}\dot{\Delta}}\\
&=\big[f+bm\,u-bm\,d(u)-x^{(n)}_d+\mathbf{\bar{c}^\mathrm{T}\tilde{x}}\big]\varepsilon
+bm\varphi^{-1}\mathbf{\Delta^\mathrm{T}\dot{\Delta}}\\
\end{align*}

\noindent
Applying the proposed control law (\ref{eq:law}) and noting that $\mathbf{\dot{\Delta}}=\mathbf{\dot{\hat{D}}}$, 
then

\begin{align*}
\dot{V}(t)&=\big[bm(\hat{d}-d)-\kappa\varepsilon\big]\varepsilon+bm\varphi^{-1}\mathbf{\Delta^\mathrm{T}
\dot{\hat{D}}}\\
&=\big[bm\mathbf{\Delta^\mathrm{T}\Psi}(\hat{u})-\kappa\varepsilon\big]\varepsilon+bm\varphi^{-1}\mathbf{
\Delta^\mathrm{T}\dot{\hat{D}}}\\
&=-\kappa\varepsilon^2+bm\varphi^{-1}\mathbf{\Delta}^\mathrm{T}\big[\mathbf{\dot{\hat{D}}}+\varphi\varepsilon
\mathbf{\Psi}(\hat{u})\big]\\
\end{align*}

\noindent
Furthermore, defining $\mathbf{\dot{\hat{D}}}$ according to (\ref{eq:adapta}), $\dot{V}(t)$ becomes

\begin{equation}
\dot{V}(t)=-\kappa\varepsilon^2
\label{eq:liap_p}
\end{equation}

\noindent
which implies that $V(t)\le V(0)$ and that $\varepsilon$ and $\mathbf{\Delta}$ are bounded. From the 
definition of $\varepsilon$ and considering Assumption~\ref{hp:traj}, it can be easily verified that 
$\dot{\varepsilon}$ is also bounded.

\noindent
To establish the convergence of the combined tracking error measure, the time derivative of $\dot{V}$ 
must be also analyzed:

\begin{equation}
\ddot{V}(t)=-2\kappa\varepsilon\dot{\varepsilon}
\label{eq:liap_pp}
\end{equation}

\noindent
which implies that $\dot{V}(t)$ is also bounded and, from Barbalat's lemma, that $\varepsilon\to0$ as 
$t\to\infty$. From the definition of limit, it means that for every $\xi>0$ there is a corresponding 
number $\tau$ such that $|\varepsilon|<\xi$ whenever $t>\tau$. According to Eq.~(\ref{eq:measu}) and 
considering that $|\varepsilon|<\xi$ may be rewritten as $-\xi<\varepsilon<\xi$, one has 

\begin{equation}
-\xi<c_0\tilde{x}^{(n-1)}+c_1\lambda\tilde{x}^{(n-2)}+\cdots+c_{n-2}\lambda^{n-2}\dot{\tilde{x}}
+c_{n-1}\lambda^{n-1}\tilde{x}<\xi
\label{eq:ebounds}
\end{equation}

\noindent
Multiplying (\ref{eq:ebounds}) by $e^{\lambda t}$ yields

\begin{equation}
-\xi e^{\lambda t}<\frac{d^{n-1}}{dt^{n-1}}(\tilde{x}e^{\lambda t})<\xi e^{\lambda t}\\ 
\label{eq:mbounds}
\end{equation}

\noindent
Thus, integrating (\ref{eq:mbounds}) $n-1$ times between $0$ and $t$ gives

\begin{multline}
-\frac{\xi}{\lambda^{n-1}}e^{\lambda t}+\left(\frac{d^{n-2}}{dt^{n-2}}(\tilde{x}e^{\lambda t})
\Big|_{t=0}+\frac{\xi}{\lambda}\right)\frac{t^{n-2}}{(n-2)!}+\cdots+\left(\tilde{x}(0)+
\frac{\xi}{\lambda^{n-1}}\right)\le\tilde{x}e^{\lambda t}\le\frac{\xi}{\lambda^{n-1}}e^{\lambda 
t}+\\+\left(\frac{d^{n-2}}{dt^{n-2}}(\tilde{x}e^{\lambda t})\Big|_{t=0}-\frac{\xi}{\lambda}\right)
\frac{t^{n-2}}{(n-2)!}+\cdots+\left(\tilde{x}(0)-\frac{\xi}{\lambda^{n-1}}\right)
\label{eq:int_n-1}
\end{multline}

\noindent
Furthermore, dividing (\ref{eq:int_n-1}) by $e^{\lambda t}$, it can be easily verified that the values of 
$\tilde{x}$ can be made arbitrarily close to $0$ (within a distance $\xi$) by taking $t$ sufficiently large 
(larger than $\tau$), i.e., $\tilde{x}\to0$ as $t\to\infty$. Considering the $(n-2)^\mathrm{th}$ integral 
of (\ref{eq:mbounds}), dividing again by $e^{\lambda t}$ and noting that $\tilde{x}$ converges to zero, 
it follows that $\dot{\tilde{x}}\to0$ as $t\to\infty$. The same procedure can be successively repeated 
until the convergence of each component of the tracking error vector is achieved: $\mathbf{\tilde{x}\to0}$ 
as $t\to\infty$.\hfill$\square$ \vspace*{12pt}

\section{Illustrative example}

In order to illustrate the controller design method and to demonstrate its performance, consider a 
forced Van der Pol oscillator

\begin{equation}
\ddot{x}-\mu(1-x^2)\dot{x}+x=b\upsilon
\label{eq:vanderpol}
\end{equation}

\begin{figure}[htb]
\centering
\hspace*{90pt}\includegraphics[width=0.7\textwidth]{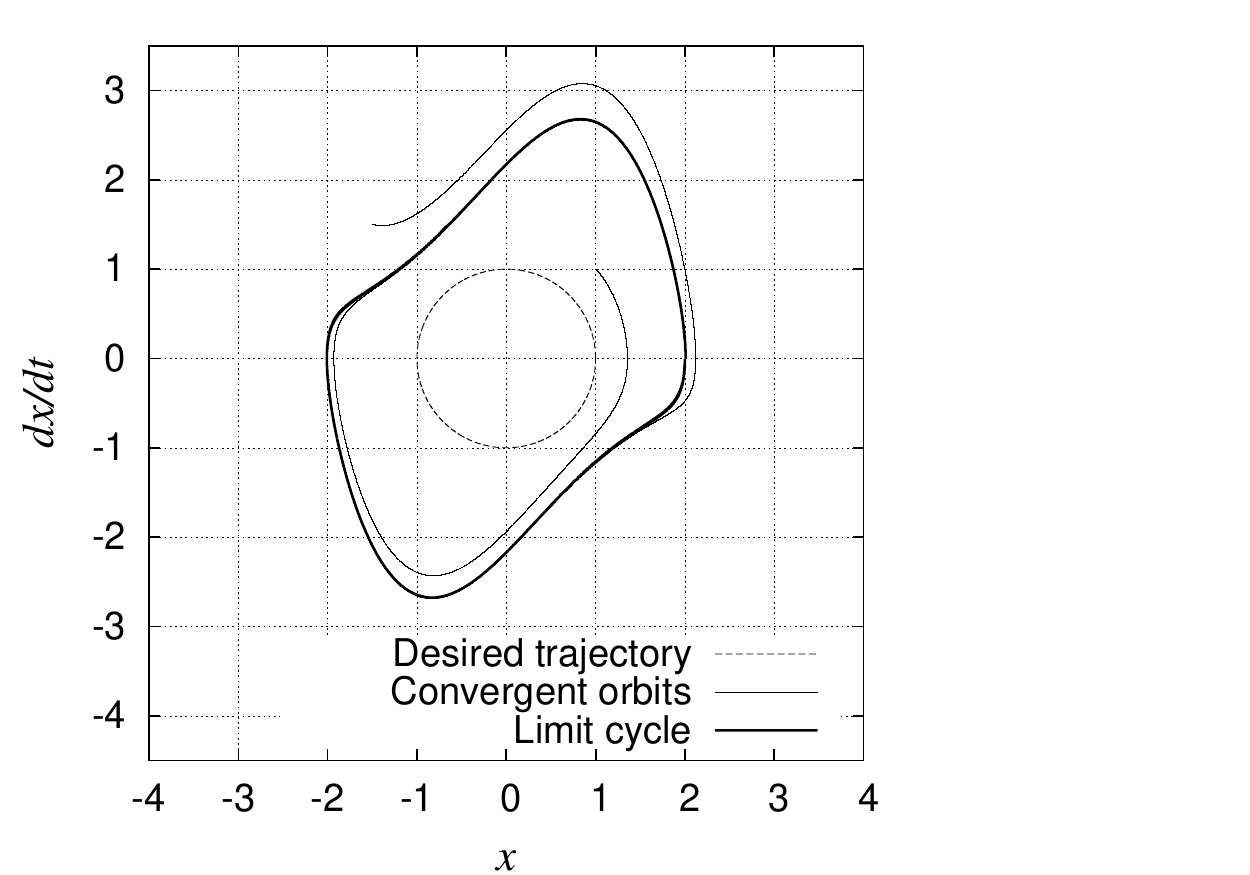}
\caption{Phase portrait of the unforced Van der Pol oscillator} 
\label{fi:vdp}
\end{figure}

Without control, i.e., by considering $\upsilon=0$, the Van der Pol oscillator exhibits a limit 
cycle. The control objective is to let the state vector $\mathbf{x}=[x,\dot{x}]$ track a desired 
trajectory $\mathbf{x}_d=[\sin t,\cos t]$ situated inside the limit cycle. Figure~\ref{fi:vdp} 
shows the phase portrait of the unforced Van der Pol oscillator with the limit cycle, two 
convergent orbits and the desired trajectory.

According to the previously described scheme and considering $\varepsilon=\dot{\tilde{x}}+\lambda
\tilde{x}$, the control law can be chosen as follows

\begin{displaymath}
u=\frac{1}{bm}[-\mu(1-x^2)\dot{x}+x+\ddot{x}_d-\lambda\dot{\tilde{x}}-\kappa\varepsilon]
+\hat{d}(\hat{u})
\end{displaymath}

The simulation studies were performed with an implementation in C, with sampling rates of 500 Hz 
for control system and 1 kHz for the Van der Pol oscillator, and the differential equations were 
numerically solved using the fourth order Runge-Kutta method. The chosen parameters were $b=1$, 
$m=1$, $\mu=1$, $\delta_l=-0.4$, $\delta_r=0.3$, $\lambda=0.6$, $\kappa=10$ and $\varphi=3$. 
Concerning the fuzzy inference system, triangular and trapezoidal membership functions, 
respectively $\mu_\mathrm{tri}$ and $\mu_\mathrm{tri}$, were adopted for $\hat{U}_r$:

\begin{equation}
\mu_\mathrm{tri}=\mathrm{max}\left\{\mathrm{min}\left(\frac{\hat{u}-a}{b-a},\frac{c-\hat{u}}{c-b}
\right),0\right\}
\label{eq:mftri}
\end{equation}

\noindent
where $a$, $b$ and $c$, with $a<b<c$, represent the abscissae of the three corners of the underlying 
triangle membership function,

\begin{equation}
\mu_\mathrm{trap}=\mathrm{max}\left\{\mathrm{min}\left(\frac{\hat{u}-a}{b-a},1,\frac{d-\hat{u}}{d-c}
\right),0\right\}
\label{eq:mftrap}
\end{equation}

\noindent
where $a$, $b$, $c$ and $d$, with $a<b<c<d$, represent the abscissae of the four corners of the 
underlying trapezoidal membership function.

\begin{figure}[htb]
\centering
\includegraphics[width=0.8\textwidth]{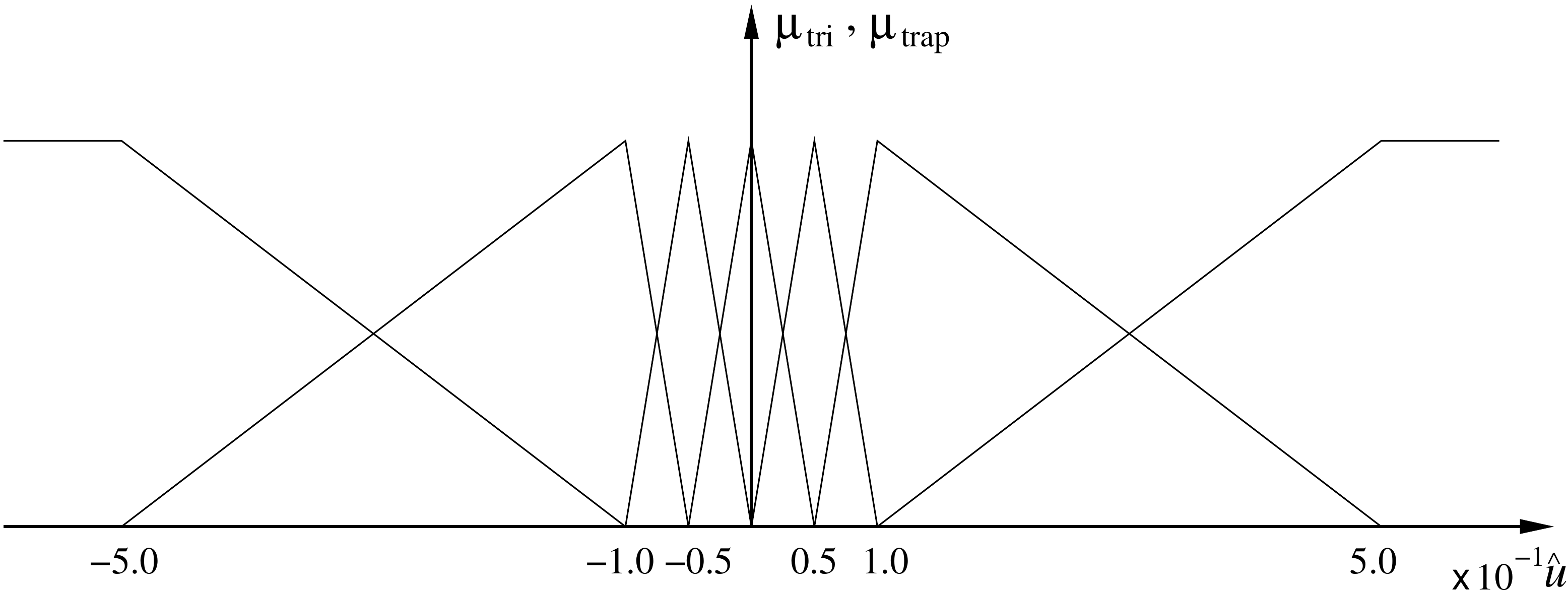}
\caption{Adopted fuzzy membership functions.} 
\label{fi:fset}
\end{figure}

The central values of the adopted membership functions were $C=\{-5.0\:;\:-1.0\:;\:-0.5\:;\:0.0\:;
\:0.5\:;\:1.0\:;\:5.0\}\times10^{-1}$ (see Fig.~\ref{fi:fset}). 

\begin{figure}[htb]
\centering
\mbox{
\subfigure[Tracking performance.]{\label{fi:graf1} 
\includegraphics[width=0.45\textwidth]{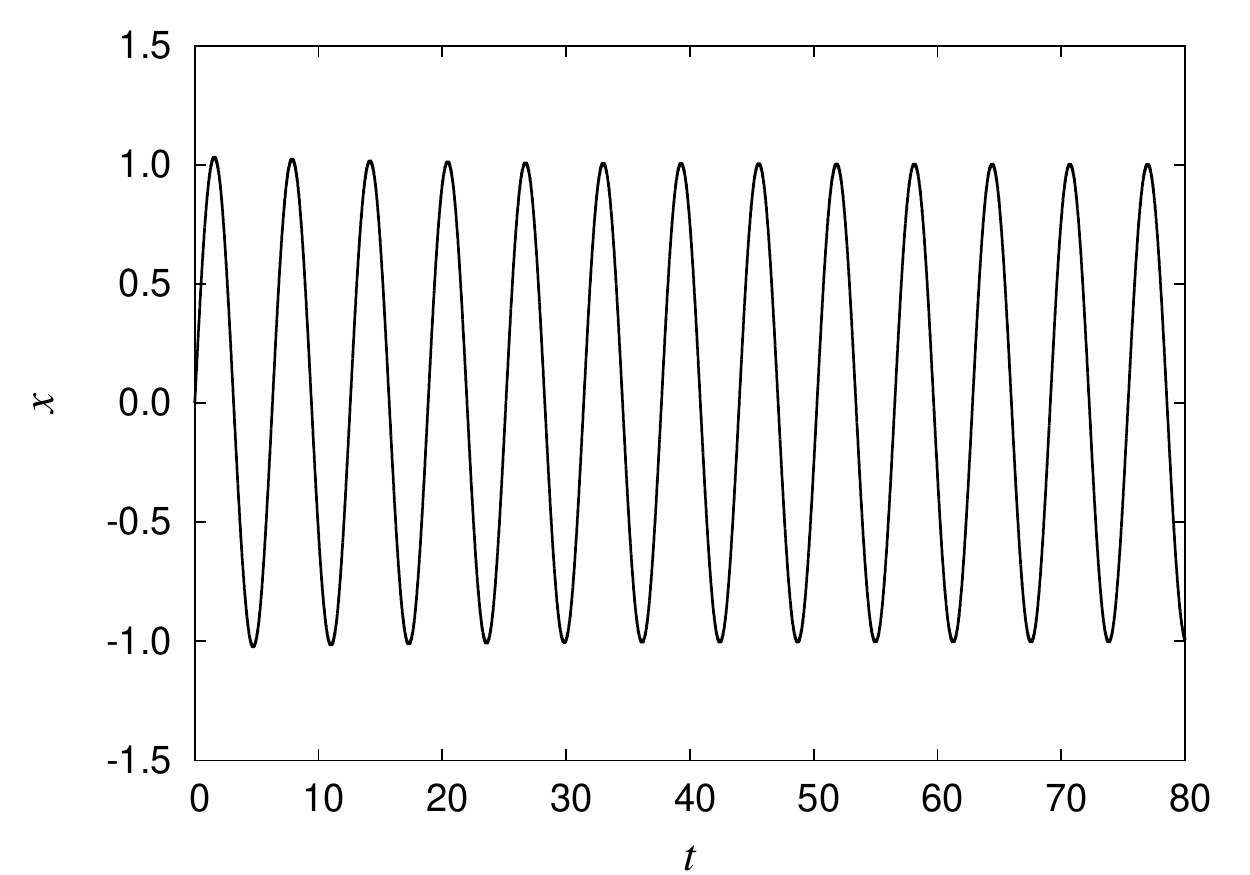}}
\subfigure[Control action.]{\label{fi:graf2} 
\includegraphics[width=0.45\textwidth]{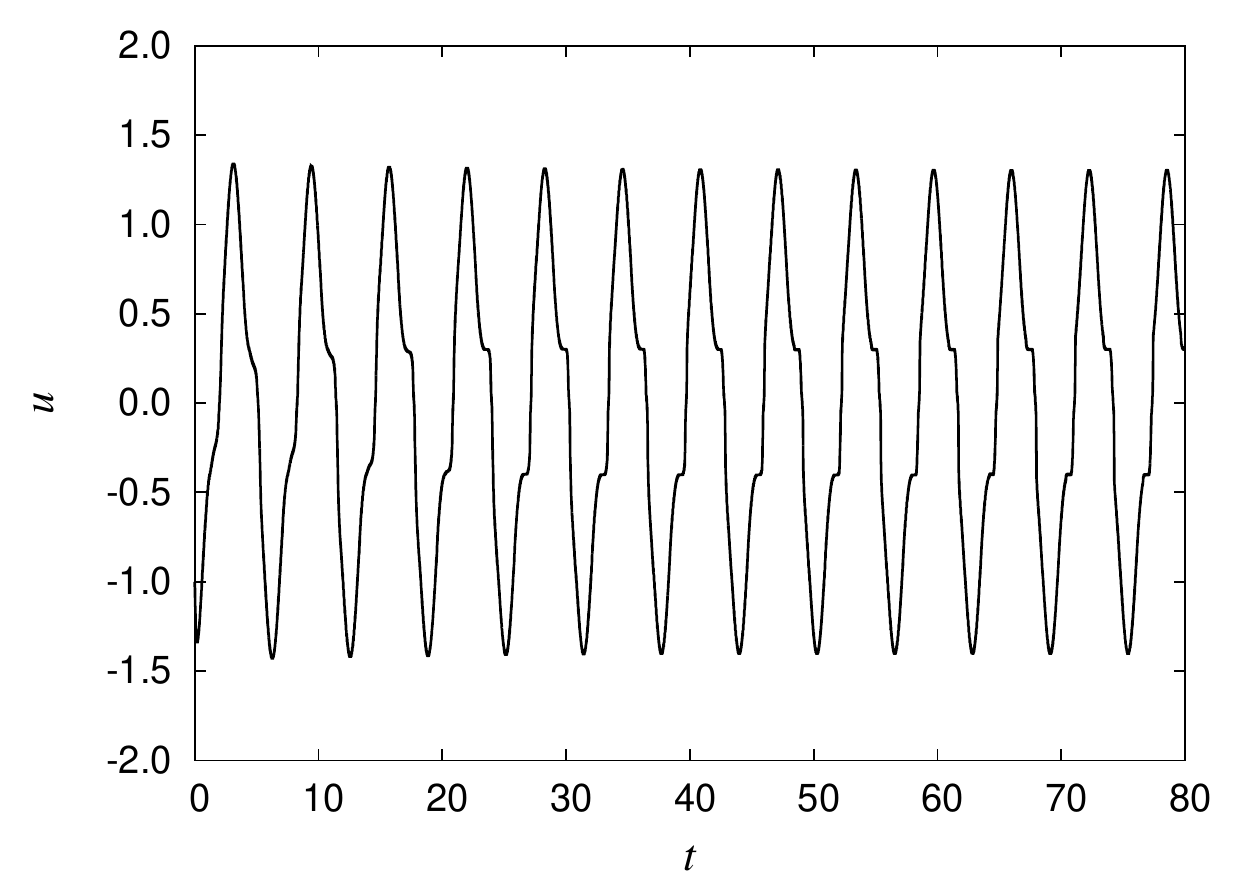}}
}
\mbox{
\subfigure[Tracking error.]{\label{fi:graf3} 
\includegraphics[width=0.45\textwidth]{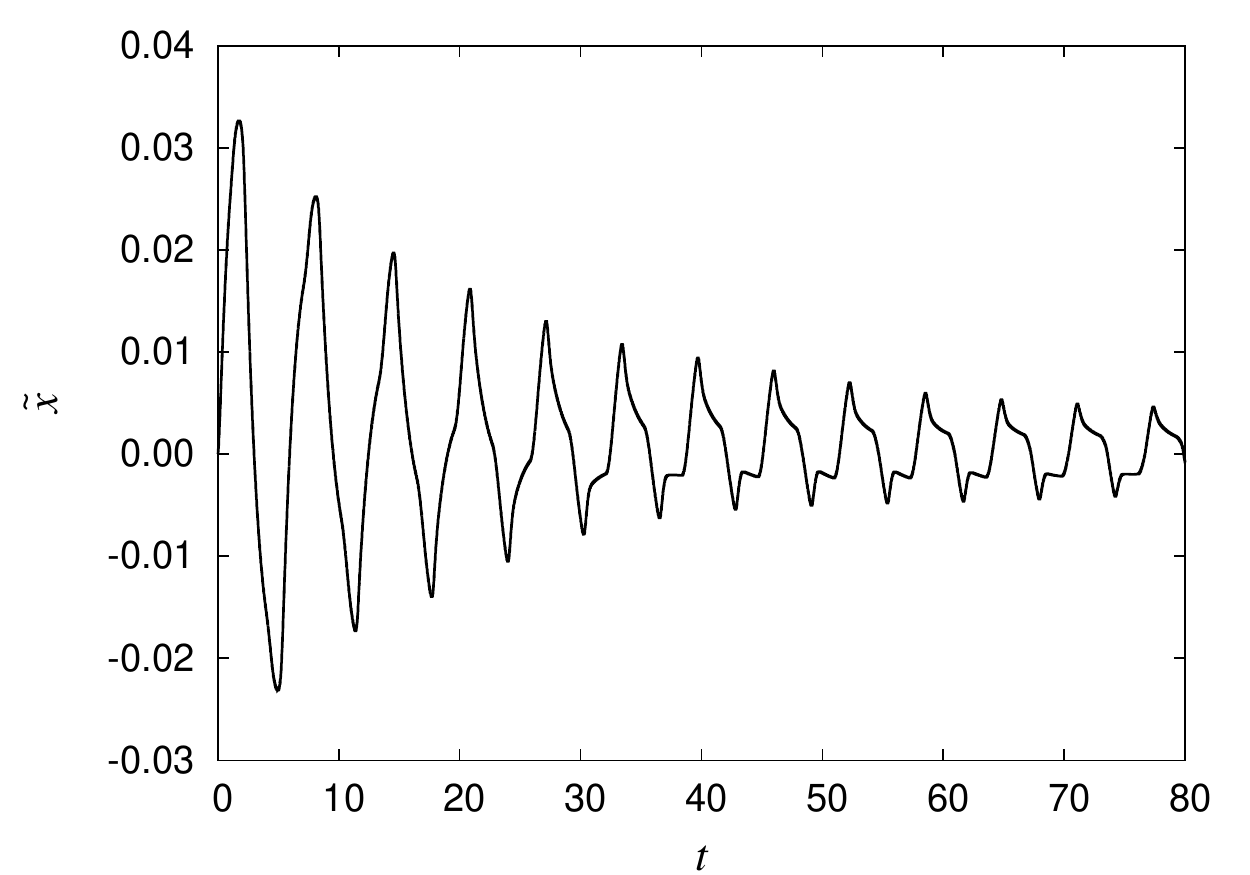}}
\subfigure[Convergence of $\hat{d}(\hat{u})$ to $d(u)$.]{\label{fi:graf4} 
\includegraphics[width=0.45\textwidth]{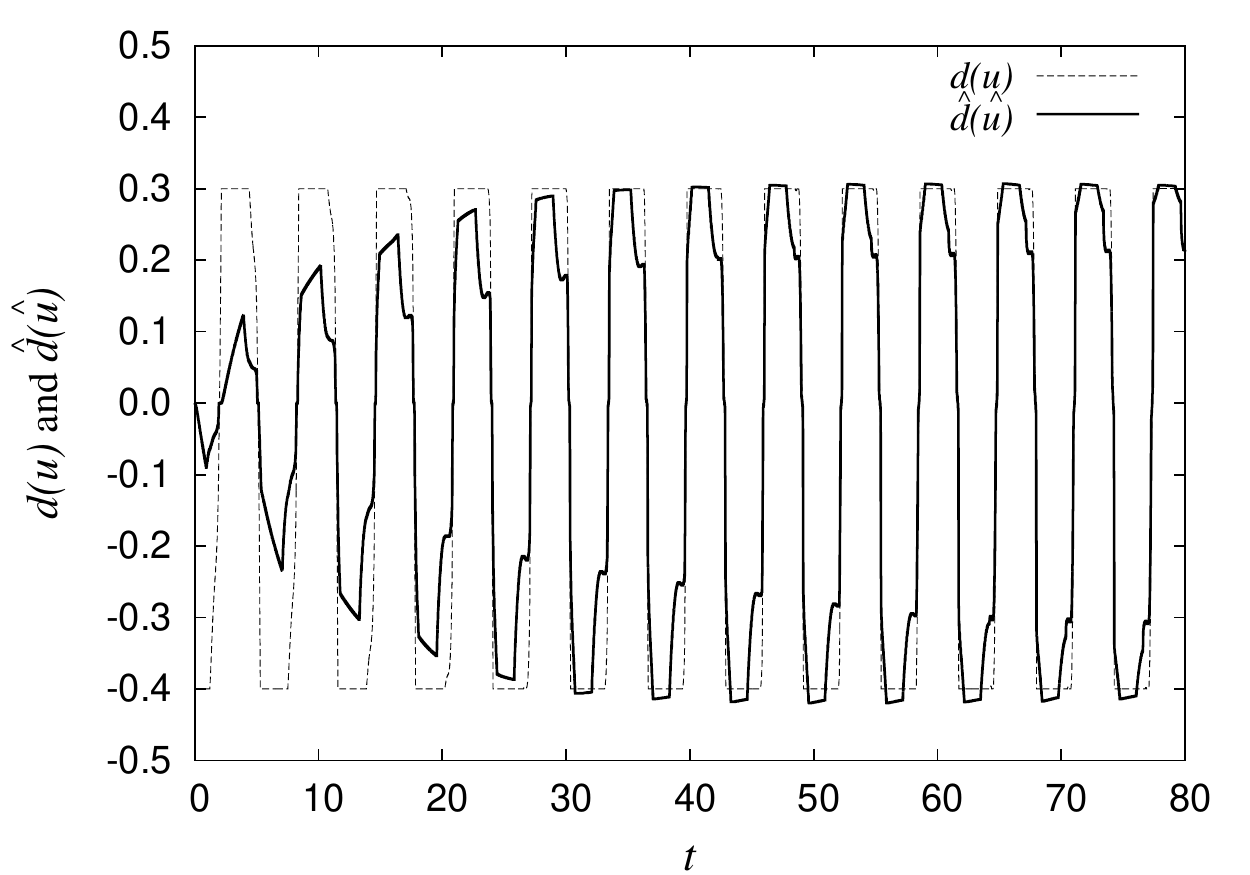}}
}
\caption{Tracking performance with $x_d=\sin t$.}
\label{fi:sim}
\end{figure}

It is also important to emphasize, that the vector of adjustable parameters was initialized with 
zero values, $\mathbf{\hat{D}=0}$, and updated at each iteration step according to the adaptation 
law presented in Eq.~(\ref{eq:adapta}). Figure~\ref{fi:sim} gives the corresponding results for the 
tracking of $x_d=\sin t$.

As observed in Fig.~\ref{fi:sim}, the proposed control law is able to provide trajectory tracking,
Fig.~\ref{fi:graf1}, with a small associated error, Fig.~\ref{fi:graf3}. Figure~\ref{fi:graf4}
shows the ability of the adaptive fuzzy scheme to recognize and previously compensate the dead-band
characteristics.

\section{CONCLUDING REMARKS}

The present work addressed the problem of controlling nonlinear systems subject to dead-zone input. An 
adaptive fuzzy controller was proposed to deal with the trajectory tracking problem. The boundedness
and convergence properties of the closed-loop signals were analytically proven using Lyapunov stability 
theory and Barbalat's lemma. The control system performance was also confirmed by means of numerical 
simulations with an application to the forced Van der Pol equation. The adaptive algorithm could 
automatically recognize the dead-zone nonlinearity and previously compensate its undesirable effects.

\section{ACKNOWLEDGEMENTS}

The authors acknowledge the support of the State of Rio de Janeiro Research Foundation (FAPERJ).

\end{document}